\RequirePackage{ifpdf}
\documentclass[preprint,notoc]{JHEP3}
\usepackage{epsfig}
\usepackage{slashed}

\def\sr2{\sqrt{2}}

\def\to{\rightarrow}

\def\bi{\begin{itemize}}
\def\ei{\end{itemize}}

\def\c1p{C1^\prime}
\def\ta{\tilde a}

\def\tG{\widetilde G}

\def\tu{\tilde u}

\def\ta{\tilde a}

\def\tst{\tilde t}

\def\tg{\tilde g}

\def\tq{\tilde q}

\def\tz{\widetilde Z}

\def\alt{\lesssim}
\def\agt{\gtrsim}
\def\be{\begin{equation}}
\def\ee{\end{equation}}
\def\bea{\begin{eqnarray}}
\def\eea{\end{eqnarray}}

\newcommand\Drv[2]{\frac{d #1}{d #2}}

\preprint{\vbox{OU-HEP-140531}}

\title{Coupled Boltzmann computation of mixed 
axion neutralino dark matter in the SUSY DFSZ axion model
}
\author{Kyu Jung Bae$^{a}$, Howard Baer$^{a}$, Andre Lessa$^{b}$ and Hasan Serce$^{a}$\\
$^a$Dept.\ of Physics and Astronomy, University of Oklahoma, Norman, OK 73019, USA\\
$^b$ Instituto de F\'isica, Universidade de S\~ao Paulo, S\~ao Paulo - SP, Brazil\\
E-mail: \email{bae@nhn.ou.edu}, \email{baer@nhn.ou.edu}, \email{lessa@fma.if.usp.br}, \email{serce@ou.edu}
}

\abstract{The supersymmetrized DFSZ axion model is highly motivated not only because it 
offers solutions to both the gauge hierarchy and strong CP problems, but also because it provides
a solution to the SUSY $\mu$-problem which naturally allows for a Little Hierarchy. 
We compute the expected mixed axion-neutralino dark matter abundance 
for the SUSY DFSZ axion model in two benchmark cases--
a natural SUSY model with a standard neutralino underabundance (SUA) and an mSUGRA/CMSSM model 
with a standard overabundance (SOA).
Our computation implements coupled Boltzmann equations which track the radiation density along with neutralino, axion, axion CO (produced via coherent oscillations), saxion, saxion CO, axino and gravitino densities.
In the SUSY DFSZ model, axions, axinos and saxions go through the process of freeze-in--
in contrast to freeze-out or out-of-equilibrium production as in the SUSY KSVZ model--
resulting in thermal yields which are largely independent of the re-heat temperature.
We find the SUA case with suppressed saxion-axion couplings ($\xi=0$) only admits solutions 
for PQ breaking scale $f_a\alt 6\times 10^{12}$ GeV 
where the bulk of parameter space tends to be axion-dominated. 
For SUA with allowed saxion-axion couplings ($\xi =1$), then $f_a$ values 
up to $\sim 10^{14}$ GeV are allowed.
For the SOA case, almost all of SUSY DFSZ parameter space is disallowed by a
combination of overproduction of dark matter, overproduction of dark radiation or violation of BBN constraints.
An exception occurs at very large $f_a\sim 10^{15}-10^{16}$ GeV where large entropy dilution from CO-produced saxions leads to allowed models.
}

\begin{document}

\section{Introduction}
\label{sec:intro}

The recent discovery of the Higgs boson~\cite{atlas_h,cms_h} with mass $m_h=125.5\pm 0.5$ GeV confirms 
the particle content of the Standard Model (SM) but carries with it a puzzle: 
why is the Higgs mass so light? 
Radiative corrections to the Higgs mass are of the form 
\be 
\delta m_h^2\sim \frac{c_i}{16\pi^2}\Lambda^2 \label{eq:dmh}
\ee 
where $c_i$ is a loop dependent factor with $|c_i|\sim 1$ and $\Lambda$ is the cutoff scale below which the SM ought to be valid. 
Setting $\delta m_h^2=m_h^2$ and using {\it e.g.} $c_i =1$ tells us that $\Lambda\alt 1$ TeV, {\it i.e.} that we
expect new physics starting near the TeV scale. Yet so far, LHC data are in strong agreement with the SM.

The introduction of supersymmetry (SUSY) into the theory tames the quadratic divergences, and furthermore
relates the Higgs mass to the $Z$ mass, predicting $m_h\alt 135$ GeV within the 
Minimal Supersymmetric Standard model, or MSSM~\cite{mhiggs}. Comparing the measured value of $m_h$ to the 
theory prediction, one finds that the Higgs mass falls squarely within the narrow window predicted by SUSY.

A further problem with the SM arises in the QCD sector, where naively the $U(2)_L\times U(2)_R$ chiral
symmetry of the light quark sector implies the existence of four-- and not three-- light pions.
't Hooft resolved this problem~\cite{tHooft} via discovery of the QCD $\theta$ vacuum where the 
anticipated $U(1)_A$ symmetry is not respected~\cite{weinberg}. 
A consequence of 't Hooft's solution is that the QCD Lagrangian contains a $CP$-violating term
\be
{\cal L}\ni \bar{\theta}\frac{g_s^2}{32\pi}F_{A\mu\nu}\tilde{F}^{\mu\nu}_A
\ee
where $\bar{\theta}\equiv\theta+arg\ det (M)$, with $M$ being the quark mass matrix.
Measurements of the neutron EDM imply $\bar{\theta}\alt 10^{-10}$, so the term is somehow minuscule.
An elegant resolution of this ``strong $CP$'' problem involves the introduction of
a spontaneously broken global Peccei-Quinn (PQ) symmetry~\cite{pq} 
and its concommitant axion field $a$~\cite{ww}. For realistic models~\cite{ksvz,dfsz}, 
the scale of PQ symmetry breaking $f_a$ is required to be $f_a\agt 10^9$ GeV lest red giant stars 
cool too quickly~\cite{axreview}.

By enlarging the SM to include the PQ axion, Eq.~\ref{eq:dmh} implies a
Higgs mass $m_h\sim f_a$. To solve the strong $CP$ problem while
simultaneously taming the Higgs mass, it seems both SUSY and PQ are required. 
In this case, the axion comprises but one element of an axion superfield given by

\be
A =\frac{s+ia}{\sqrt{2}}+\sqrt{2}\theta \ta +\theta^2{\cal F}_a
\ee
where now the $\theta$ are spinorial Grassmann co-ordinates and ${\cal F}_a$ is the
axion auxiliary field. Here, $s$ is the $R$-parity even spin-0 {\it saxion} field and $\ta$
is the $R$-parity-odd spin-$1\over 2$ {\it axino} field.\footnote{
It is worth noting that we describe the axion superfield below the PQ symmetry breaking scale, so it is non-linearly realized with a superpotential given by
$W=\mu e^{c_H A/v_{PQ}}H_uH_d$ where $c_H$ is the PQ charge of the Higgs superfield bilinear 
and $v_{PQ}$ denotes the vev from PQ symetry breaking.
The axion superfield transforms under the PQ symmetry as
$A\to A+i\alpha v_{PQ}$ while the Higgs fields transform as $H_uH_d\to e^{-ic_H\alpha}$
where $\alpha$ is an arbitrary real number. 
The SUSY DFSZ axion model respects this shift symmetry unless we consider the chiral symmetry 
breaking that produces the axion potential.
In comparing our notation against Ref.~\cite{Dreiner14}, 
what we call $A$ is denoted there as $\hat{\Phi}_a$.}
In gravity-mediated SUSY breaking models, then $m_s$ is a soft SUSY breaking term which is
expected to be $\sim m_{3/2}$ and the (more model-dependent) axino mass
$m_{\ta}$ is also expected to be of order $m_{3/2}$~\cite{cl}.
Here, the gravitino mass $m_{3/2}$ generated via the super-Higgs mechanism is expected to be of order 
the weak scale $\sim 1$ TeV while the visible sector sparticle masses are also expected to be of 
order $m_{3/2}$~\cite{sw}.
Lack of a SUSY signal at LHC8, along with a decoupling solution~\cite{dine} to the SUSY flavor, $CP$, 
proton decay and gravitino problems all suggest $m_{3/2}$ to be more like $\sim 10-20$ TeV. 
Meanwhile, SUSY electroweak naturalness requires the superpotential $\mu$-term to be 
$\sim 100-200$ GeV~\cite{bbmp}.
In such a case, one would expect the lightest neutralino to be the stable LSP and to be a 
higgsino-like WIMP dark matter candidate. 
However, in this case dark matter would be composed of an axion-neutralino admixture, 
{\it i.e.} {\it two dark matter particles}! 

A further problem with SUSY models is the so-called $\mu$-problem. 
The superpotential Higgs/higgsino mass term $\mu$ is supersymmetric so that one expects it 
naively to have values of order the GUT or reduced Planck scales. 
But since it gives mass to the newly discovered Higgs boson (along with $W^\pm$ and $Z^0$), 
 phenomenology dictates it to be of order the weak scale.
An elegant solution occurs within the context
of the SUSY DFSZ axion model~\cite{dfsz}. In this case, the SM Higgs doublets carry PQ charge so that the 
$\mu$ term is in fact forbidden. But there may exist non-renormalizable couplings of the Higgs doublets 
to a PQ-charged superfield $S$:
\be
W_{\rm DFSZ}\ni \lambda \frac{S^{n+1}}{M_P^n} H_u H_d 
\label{eq:Wdfsz}
\ee
where $n$ is an integer $\ge 1$.
In this Kim-Nilles solution to the SUSY $\mu$ problem~\cite{kimnilles}, 
under PQ symmetry breaking $S$ receives a vev $\langle S\rangle\sim f_a$ so that an effective $\mu$ term is 
generated with 
\be
\mu\sim \lambda f_a^{n+1}/M_P^n .
\ee
This mechanism allows for $\mu \ll m_{3/2}$ since the $\mu$-term arises from PQ symmetry breaking whilst 
$m_{3/2}$ might arise from hidden sector SUSY breaking.\footnote{Historically, Kim-Nilles sought
to relate $\mu\sim m_{3/2}$ in this approach.}
For $n=1$ and $\lambda \sim 1$, $\mu\sim 100$ GeV  requires
$f_a\sim 10^{10}$ GeV while $n > 1$ allows for much larger values of $f_a$.
Alternatively, the Giudice-Masiero solution~\cite{gm} to the $\mu$-problem
favors $\mu\sim m_{3/2}$ wherein tension then arises between SUSY naturalness
and LHC sparticle mass bounds.

In the supersymmetric DFSZ model, the axion domain wall number
is $N_{DW} = 6$ since the quark doublet superfields carry PQ charge. As a result, the
PQ symmetry must be broken before or during inflation\footnote{This usually
imposes an upper bound on the re-heat temperature $T_R$. However, as discussed
below, in the DFSZ scenario the thermal production of axions, saxions and axinos is independent of $T_R$.}
in order to avoid the overclosure of the universe through the production of
stable domain walls~\cite{Sikivie:1982qv}. 
In this case the axion misalignment angle ($\theta_i$) is constant in
our patch of the universe and the relic density from coherent
oscillations of the axion is given by (ignoring possible entropy dilution effects)
~\cite{vacmis,bhk08,vg1}:
\be
\Omega_a^{\rm std} h^2 \simeq 0.23
f(\theta_i)\theta_i^2\left(\frac{f_a/N_{DW}}{10^{12} \mbox{ GeV}}\right)^{7/6}
\label{eq:omegaA}
\ee
where $f(\theta_i) = \left[\ln\left(e/(1-\theta_i^2/\pi^2\right)\right]^{7/6}$.

It has been pointed out recently~\cite{axionBICEP2,axionBICEP3,axionBICEP1} that
the measurement of a large tensor-to-scalar ratio ($r \simeq 0.2$) by the BICEP2
collaboration~\cite{BICEP} provides strong
constraints on axion models. In particular, the breaking of the PQ symmetry {\it before} inflation 
(as required in the DFSZ model) would lead to too large isocurvature perturbations 
thus excluding this possibility. 
However, simple extensions of the PQ breaking sector are possible that can significantly
affect the inflationary cosmology.
One possible extension is to introduce an inflaton-dependent interaction that explicitly breaks the PQ symmetry.
In this case the axion becomes massive during inflation and isocurvature
perturbations do not develop~\cite{axionBICEP2}.
Another possibility is to consider the case where the PQ breaking scale during
inflation is larger than in the current universe, so isocurvature
perturbations are suppressed.
This scenario can be realized through the $D$-term
interaction of the anomalous $U(1)$ gauge symmetry in the PQ sector~\cite{cjsBICEP} or 
from Planck-suppressed interactions between the axion superfield and the 
inflaton superfield in the K\"ahler potential~\cite{ejcBICEP}.
In the following discussion, 
we may assume that the isocurvature perturbation is suppressed by one of these extended 
PQ breaking scenarios, so the SUSY DFSZ model can be made compatible with the BICEP2 measurement.
Alternatively, it remains to be seen whether the BICEP2 result is verified by further
measurements at different frequency values\cite{seljak,spergel}.

Besides being produced through coherent oscillations, axions are also produced
through thermal scatterings in the early universe. In this case, however, they
are relativistic and constitute dark radiation, contributing to the number of
relativistic degrees of freedom in the early universe. 
The amount of dark radiation produced during big bang nucleosynthesis (BBN) or during
matter-radiation decoupling is usually parametrized by the number of
effective neutrinos, which is conservatively constrained by BBN and CMB data to be 
$N_{\rm eff} < 4.6$  (or $\Delta N_{\rm eff} < 1.6$).\footnote{The Planck 
experiment\cite{Ade:2013zuv} has
recently published $N_{eff}=3.30\pm 0.27$ in apparent agreement with the SM prediction.}
 However, as discussed
in Sec.~\ref{sec:boltz}, the thermal production (TP-) of axions is suppressed at temperatures below the
Higgs/higgsino masses, resulting in a negligible contribution to $\Delta
N_{\rm eff}$. Nonetheless, relativistic axions may also be produced from saxion
decays. The $s \to aa$ branching ratio is controlled by the axion-saxion
effective coupling~\cite{cl}:
\be
\mathcal{L} \ni \frac{\xi}{f_a} s \left[ \left(\partial_\mu a \right)^2 
+ i \bar{\ta} \slashed \partial \ta \right] \label{eq:xicoup}
\ee
where $\xi$ is a model dependent parameter, which can be small (or even zero) or
as large as 1. Since the saxion decays strongly depend on $\xi$, 
we discuss the $\xi = 0$ and $\xi =1$ limiting cases separately in Sec.~\ref{sec:results}.

As already mentioned above, the total DM abundance in the SUSY DFSZ scenario
receives contribution from both CO axions and relic neutralinos. The relic
abundance of neutralinos in the SUSY DFSZ model was first considered in 
Ref.~\cite{chun,bci,ltr,dfsz1}.
Neutralinos are produced through the usual freeze-out mechanism as well as
through injection from saxion and axino decays. Therefore, in order to compute
the final neutralino relic abundance it is necessary to determine the axino and
saxion production rates and decay widths in the early universe. The axion
multiplet couples to the MSSM primarily through its coupling with the Higgs
supermultiplets, generated after breaking of the PQ symmetry as~\cite{dfsz1}:
\be
{\cal L}_{{\rm DFSZ}}=\int d^2\theta (1+B\theta^2)\mu e^{c_H  A/v_{PQ}}  H_u  H_d,
\label{eq:superptl}
\ee
where $1+B \theta^2$ is a SUSY breaking spurion field.
Since Eq.~\ref{eq:superptl} generates tree level interactions of the type $A H_u
H_d$, the thermal production of saxions, axions and axinos happen through the
freeze-in mechanism~\cite{lhall}.
In this case the production is maximal
at $T \sim m_{s,\tilde{a}}$, leading to thermal yields which are largely {\it independent}
of the re-heat temperature ($T_R$)~\cite{bci,bci11}. As discussed in
Sec.~\ref{sec:boltz}, in some regions of parameter space the thermal production
and decay of axinos and saxions are competing processes and cannot
be treated separately. As a result, the sudden decay approximation is no
longer valid and a precise calculation of the neutralino relic abundance (which
receives contributions from axino and saxion decays) requires the numerical
integration of the Boltzmann equations.

In the present work, we continue to refine the calculation of mixed
axion-neutralino CDM in the SUSY DFSZ model. Here we compute the evolution of the axion, axino,
saxion, neutralino and gravitino relic abundances using the appropriate system
of coupled Boltzmann equations.
In Ref's.~\cite{bls,bbl}, a similar calculation was performed in the SUSY KSVZ
scenario that allowed for a more precise computation of the dark matter relic
abundance; this method included the effects of the temperature-dependence of the
neutralino annihilation cross section ($\langle\sigma v\rangle (T)$) and the
non-thermal production of neutralinos in models with large entropy injection
from saxion decays.
Here we apply a similar formalism to the SUSY DFSZ model, using the 
axino/saxion thermal production rates and decay rates computed in previous
works~\cite{bci,dfsz1,bci11}. This approach allows for
\begin{itemize}
\item correct calculation of axino and saxion thermal yields for small $f_a$
values,
\item inclusion of temperature-dependent $\langle\sigma v\rangle (T)$ such as occurs
for bino-like CDM with mainly $p$-wave annihilation,
\item inclusion of non-sudden axino/saxion decays and
\item accurate calculation of entropy production and injection in the early
universe.
\end{itemize}
Furthermore, we also scan the parameter space of the SUSY DFSZ model
and identify the regions consistent with dark matter, BBN and dark radiation
constraints.

The remainder of this paper is organized as follows. In Sec.~\ref{sec:boltz}, 
we discuss the set of coupled Boltzmann equations used to compute our numerical results. 
In Sec.~\ref{sec:results}, we present the two benchmark models used in our analysis
and discuss the behavior of the dark matter relic abundance in these models
as a function of the PQ parameters. In order to keep our results general, we
scan over the most relevant PQ parameters and numerically solve the Boltzmann
equations for each point. We also discuss the BBN and $\Delta N_{\rm eff}$
constraints in these models.
Finally, in Sec.~\ref{sec:conclude}, we present a brief summary and conclusions.

\section{Coupled Boltzmann equations}
\label{sec:boltz}

Our goal is to numerically solve the coupled Boltzmann equations which track the 
number and energy densities of neutralinos $\tz_1$, gravitinos $\tG$, saxions
$s$, axinos $\ta$, axions $a$ and radiation as a function of time starting at
the re-heat temperature $T=T_R$ at the end of inflation until today.
For axions and saxions, we separately include coherent oscillating (CO) components. The simplified set of Boltzmann equations for the
SUSY KSVZ model as well as the method for their numerical solution were
presented in detail in Ref's.~\cite{bls,bbl}. In this section we discuss
the main differences between the KSVZ and DFSZ scenarios and
how the simplified Boltzmann equations derived for the KSVZ case must be generalized 
in order to allow for a proper computation of the relic abundances in the DFSZ
model.

In the KSVZ model considered in Ref.~\cite{bls}, the thermal production of
saxions, axions and axinos is maximal at $T \sim T_R$ (for re-heat
temperatures below the decoupling temperature of saxions and axinos),
resulting in a thermal yield proportional to the re-heat temperature~\cite{axprod}.
Also, since the axino/saxion decay widths are suppressed by the loop factor as well as by the PQ scale,
their decays tend to take place at temperatures $T \ll m$, where $m$ is the
axino or saxion mass.
Hence the thermal production and decay processes can be safely treated as
taking place at distinct time scales. Furthermore, the inverse decay process
($a + b \to \tilde{a},s$) is always Boltzmann-suppressed when the decay term becomes sizable 
($\Gamma\sim H$),
thus we can easily neglect the inverse decay contributions.

In the DFSZ scenario, however, the situation can be drastically different. 
Here, the tree-level couplings between the axion supermultiplet and the Higgs superfields
(Eq.~\ref{eq:superptl}) modify the thermal scatterings of saxions, 
axions and axinos and can significantly enhance their decay widths.
From the results of Ref.~\cite{bci11} (Table 1), we can 
estimate the scattering cross section (in the supersymmetric limit) by
\begin{equation}
\sigma_{(I+J\rightarrow \tilde{a}+\cdots)}(s)\sim \frac{1}{16\pi s}\left|{\cal M}\right|^2
\sim \frac{g^2c_H^2\left|T_{ij}(\Phi)^a\right|^2}{2\pi
s}\frac{M_{\Phi}^2}{v_{PQ}^2},
\end{equation}
where $\Phi$ is a PQ- and gauge-charged matter supermultiplet, $g$ the corresponding
gauge coupling constant, $T_{ij}(\Phi)^a$ is the gauge-charge matrix of $\Phi$ and $M_{\Phi}$  its
mass.
For the DFSZ SUSY axion model, 
the heaviest PQ charged superfields are the Higgs doublets, 
so $g$ is the $SU(2)$ gauge coupling, $M_{\Phi}=\mu$, and 
$\left|T_{ij}(\Phi)^a\right|^2=(N^2-1)/2=3/2$.
We can obtain the rate for the scattering contribution of axino (or saxion) production from the 
integration formula~\cite{chkl99}
\begin{equation}
\langle\sigma_{(I+J\rightarrow \tilde{a}(s)+\cdots)} v\rangle n_In_J \simeq
\frac{T^6}{16\pi^4}\int^{\infty}_{M/T}dx K_1(x) x^4 \sigma(x^2T^2)
\end{equation}
where the $K_1$ is the modified Bessel function, $M$ is the threshold energy for
the process (either the higgsino or
saxion/axino mass) and we have assumed $T \gtrsim M$.
Integrating over the Bessel function, we find that the axino (or saxion) production rate is
proportional to~\cite{bci11}:
\be
\langle\sigma_{(I+J\rightarrow \tilde{a}(s)+\cdots)} v\rangle \propto
\left(\frac{\mu}{f_a}\right)^2 
\frac{M^2}{T^4} K_2\left(M/T\right)
\label{eq:axn_prd}
\ee
where we used $n_I n_J \propto T^6$.
From the above expression (unlike the KSVZ case),  production is
maximal at $T \simeq M/3 \ll T_R$. Hence most of the thermal production of
axinos and saxions takes place at $T\sim M$, resulting in 
thermal yields which are independent of $T_R$.
This behavior is similar to the freeze-in mechanism~\cite{lhall}, where a weakly
interacting (and decoupled) dark matter particle becomes increasingly coupled
to the thermal bath as the universe cools down. However, in the
current scenario, the ``frozen-in'' species (axinos and saxions) are not stable
and their decays will only contribute to the dark matter (neutralino)
relic abundance if they take place after neutralino freeze-out 
and will also contribute to the dark radiation (axion) density.

The coupling in Eq.~\ref{eq:superptl} can also enhance the axino/saxion decay width
for large $\mu$ values, since the coupling to Higgs/higgsinos is proportional to
$\mu/f_a$. As a result, saxions and axinos may decay at much earlier times
(larger temperatures) when compared to the KSVZ scenario. If their decay
temperatures are of order of their masses, then inverse decay processes such as
$\tz_1 + h \to \tilde{a}$ or $h + h \to s$ can no longer
be neglected. In fact, in Ref.~\cite{dfsz1} it was shown that the decay
temperatures can indeed be larger than the axino or saxion mass, 
so the inverse decay process can be significant. 
The main effect of including the inverse decay process is to delay the 
axino/saxion decay. This is an important effect which cannot be
accounted for in the sudden decay approximation and requires
the numerical solution of the Boltzmann equations.
We point out, however, that the inverse decay process is only relevant for
$T_{decay} \gtrsim M$, since if the decay happens at lower temperatures, the
inverse decay process is Boltzmann-suppressed. As a result, the inverse decay
process will only be relevant for the cases where axinos and saxions decay
before neutralino freeze-out (since $T_{fr} \sim m_{\tz_1}/20 \ll M$) and we do
not expect it to modify the neutralino relic abundance. Nevertheless, it is
essential to include inverse decays in the Boltzmann equations for consistency.

As discussed above, inverse decay processes were not relevant for the KSVZ case
and were neglected in Ref's~\cite{bls,bbl}. With the
addition of the inverse decay process, the Boltzmann equations for the number ($n_i$) and energy ($\rho_i$)
densities of a thermal species $i$ ($= a,s$ or $\tilde{a}$) reads:\footnote{The
generalization of the Boltzmann equations to include decays to $n$-body final
states ($n>2$) is straightforward. For the cases where 3-body decays are
relevant (such as gravitino decays), we use the appropriate generalized
equations.}
\bea
\Drv{n_i}{t} & + & 3H n_i  =  \sum_{j\in{\rm MSSM}}\left( \bar{n}_i\bar{n}_j - n_in_j \right) \langle
\sigma v \rangle_{ij}- \Gamma_i m_i \frac{n_i}{\rho_i}\left(n_i - \bar{n}_i
\sum_{i\to a + b} \mathcal{B}_{ab} \frac{n_a n_b}{\bar{n}_a
\bar{n}_b} \right)
\nonumber
\\
& + & \sum_a 
\Gamma_a \mathcal{B}_i m_a \frac{n_a}{\rho_a} \left(n_a - \bar{n}_a \sum_{a \to
i + b} \frac{\mathcal{B}_{ib}}{\mathcal{B}_{i}} \frac{n_i n_b
}{\bar{n}_i \bar{n}_b} \right) \label{eq:nieq} \\
\Drv{\rho_i}{t} & + & 3H (\rho_i + P_i) = \sum_{j\in{\rm MSSM}}\left(  \bar{n}_i\bar{n}_j - n_in_j \right)
\langle \sigma v \rangle_{ij} \frac{\rho_i}{n_i} - \Gamma_i m_i \left( n_i -
\bar{n}_i \sum_{i\to a + b} \mathcal{B}_{ab} \frac{n_a
n_b}{\bar{n}_a \bar{n}_b}\right) \nonumber \\
 & + & \sum_a \Gamma_a  \mathcal{B}_i \frac{m_a}{2} \left( n_a -
 \bar{n}_a \sum_{a \to i + b}  \frac{\mathcal{B}_{ib}}{\mathcal{B}_{i}}
 \frac{n_i n_b}{\bar{n}_i \bar{n}_b} \right)
\label{eq:rhoieq}
\eea
where $\mathcal{B}_{ab} \equiv BR(i \to a + b)$, $\mathcal{B}_{ib} \equiv
BR(a \to i + b)$, $\mathcal{B}_i \equiv \sum_b \mathcal{B}_{ib}$,
$\bar{n}_i$ is the equilibrium density of particle species $i$ and the $\Gamma_i$ are
the zero temperature decay widths. 
The MSSM particles that interact with axion, saxion and axino are denoted by subscript $j$.
It is also convenient to use the above results 
to obtain a simpler equation for $\rho_i/n_i$:
\be
\Drv{\left(\rho_i/n_i\right)}{t} = -3 H \frac{P_i}{n_i} + \sum_{a}
\mathcal{B}_{i} \frac{\Gamma_a m_a}{n_i} \left( \frac{1}{2} 
- \frac{n_a}{\rho_a} \frac{\rho_i}{n_i} \right) \left(n_a - \bar{n}_a \sum_{a
\to i+b} \frac{\mathcal{B}_{ib}}{\mathcal{B}_{i}} \frac{n_i n_b}{\bar{n}_i
\bar{n}_b}\right) \label{eq:Rieq}
\ee
where $P_i$ is the pressure density ($P_i \simeq 0\ (\rho_i/3)$ for
non-relativistic (relativistic) particles). As discussed in Ref.~\cite{bls}, we
track separately the CO-produced components of the axion 
and saxion fields since we assume the CO components do not have scattering contributions.
Under this approximation, the equations for the CO-produced fields (axions and saxions) read:
\be
\Drv{n_i^{\rm CO}}{t} + 3H n_i^{\rm CO}  =  - \Gamma_i m_i
\frac{n_i^{\rm CO}}{\rho_i^{\rm CO}} n_i^{\rm CO} \;\;\mbox{ and }\;\;
\Drv{\left(\rho_i^{\rm CO}/n_i^{\rm CO}\right)}{t} = 0 .
\ee
The amplitude of the coherent oscillations is defined by the initial field
values, which for the case of PQ breaking before the end of inflation is a
free parameter for both the axion and saxion fields. We parametrize the initial field values by
$\theta_i = a_0/f_a$ and $\theta_s = s_0/f_a$.

Finally, we must supplement the above set of
simplified Boltzmann equations with an equation for the entropy of the thermal bath:
\be
\Drv{S}{t} = \frac{R^3}{T}\sum_i BR(i,X)
\Gamma_i m_i\left(n_i - \bar{n}_i \sum_{i\to a + b} \mathcal{B}_{ab}
\frac{n_a n_b}{\bar{n}_a
\bar{n}_b} \right) \label{Seq}
\ee
where $R$ is the scale factor and $BR(i,X)$ is the fraction of energy injected
in the thermal bath from $i$ decays.

In order to solve the above equations, it is necessary to compute the values of
the decay widths and annihilation cross sections appearing in
Eqs.~\ref{eq:nieq}, \ref{eq:Rieq} and \ref{Seq}. Since these have been presented
in previous works, we just refer the reader to the relevant references.
The MSSM particles are in thermal equilibrium in most cases, so we make a further approximation as $n_j\simeq \bar{n}_j$ in Eqs.~\ref{eq:nieq} and \ref{eq:rhoieq}.
The value of $\langle\sigma v\rangle$ for thermal axino production
is given in Ref's~\cite{bci,bci11}, while $\langle\sigma v\rangle$ for
neutralino annihilation is extracted from IsaReD~\cite{isared}.
For thermal saxion and axion production, it is reasonable to expect
annihilation/production rates similar to axino's,
since supersymmetry assures the same dimensionless couplings.
Hence we apply the result for axino thermal production from
Ref's~\cite{bci,bci11} to saxions and axions.
For the gravitino thermal production we use the result in
Ref.~\cite{psgravitino}.
The necessary saxion and axino partial widths and branching fractions can be
found in Ref.~\cite{dfsz1},
while the gravitino widths are computed in Ref.~\cite{moroigravitino}.

\begin{figure}
\begin{center}
\includegraphics[height=9cm]{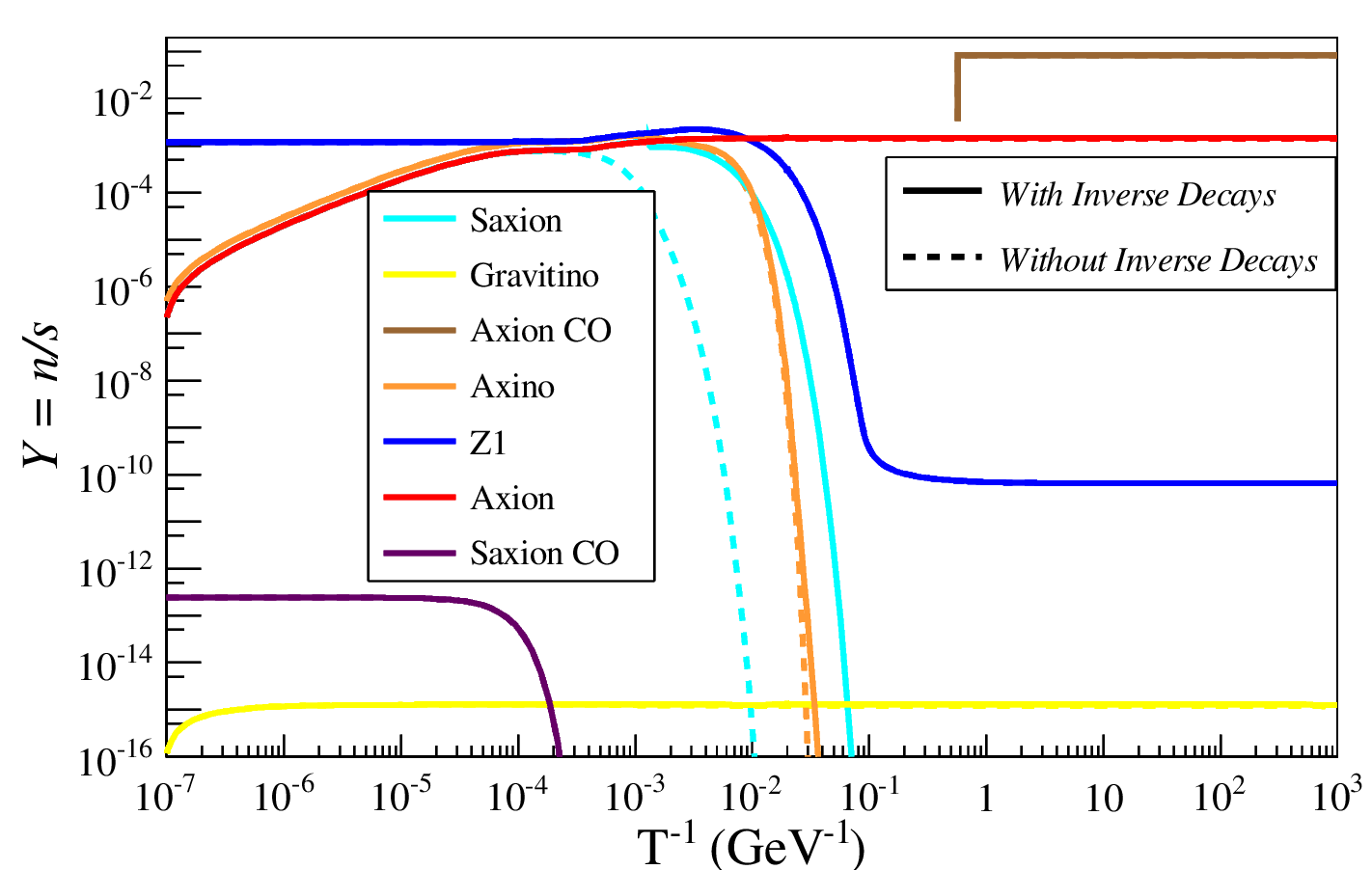}
\caption{Evolution of the axion, saxion, axino, neutralino and gravitino yields
for the SOA benchmark case with $f_a=10^{10}$ GeV, $m_{\tG}=10$ TeV, $m_{\ta}=$
1 TeV, $m_s= 500$ GeV, $\theta_s = \theta_i = 1$ and $\xi=1$.
\label{fig:yield}}
\end{center}
\end{figure}
In order to illustrate the effects discussed above, in Fig.~\ref{fig:yield}
we show a specific solution of the Boltzmann equations,
where we take a MSSM model with $\mu = 2.6$ TeV (the SOA benchmark defined in
Sec.~\ref{sec:results}) and $T_R=10^7$ GeV, $f_a=10^{10}$ GeV, $m_{\tG}=10$
TeV, $m_{\ta}=$ 1 TeV and $m_s= 500$ GeV.
We also take the saxion and axion mis-alignment angles ($\theta_s$ and
$\theta_i$) equal to 1 and $\xi =1$ so $s\to aa$ and
$s\to\ta\ta$ decays are turned on.
The figure shows the evolution of the yields ($n_i/s$) versus the inverse of the temperature. First we point out
that, as expected in the DFSZ case, saxion and axino yields (non-CO)
increase as the temperature is reduced, reaching their maximal value just before
their decay. This example clearly shows how both thermal production
and decay processes happen simultaneously, as previously discussed.
The axion follows a similar behavior, but since the axion
is (effectively) stable, its yield remains constant after the thermal production
becomes suppressed at $T \lesssim \mu$.
Gravitinos are also produced through thermal scatterings. However, as seen in
Fig.~\ref{fig:yield}, their production cross-section peaks at $T \sim T_R$, much
like the saxion/axino production in the KSVZ case. The small increase in the yields around $T = 1$ TeV
is due to the reduction in the number of relativistic SUSY degrees of freedom in
the thermal bath, which reduces the entropy density.
We also show as dashed lines the respective yields {\it without} the inclusion
of the inverse decay process. As seen in Fig.~\ref{fig:yield}, the inclusion of
inverse decays delays the decay of saxions and axinos, with the effect being
larger for saxions, since they tend to decay earlier. Nonetheless, the
neutralino and axion relic densities are unchanged, as expected from the
discussion above. For the current point chosen, the final neutralino
relic density equals its MSSM value and is well above the
experimental limits: $\Omega_{\tz_1} h^2 = 6.8$.

In the following sections, we will apply the Boltzmann equations presented here
to numerically compute the neutralino and axion relic abundances. We will also
compute the axion abundance (including the contributions from saxion
decays) in order to evaluate its contribution to the number of effective
neutrinos (dark radiation), as discussed in Sec.~\ref{sec:intro}.

\section{Numerical Results}
\label{sec:results}

\subsection{Benchmark points}

In order to compute the dark matter relic abundance in the SUSY DFSZ model
we must specify both the PQ and the MSSM parameters. Since the axion
supermultiplet interactions are proportional to $\mu$, we consider in
our numerical analysis two benchmark MSSM points: one with a small and one with
a large  value of $\mu$. In the first case (SUA), the neutralino
LSP is mostly a higgsino, resulting in a standard {\it underabundance} of
neutralino cold dark matter (CDM). The second benchmark, which we label SOA, 
has a standard thermal {\it overabundance} of neutralino dark matter,
since the neutralino is mostly a bino.

%
\begin{table}\centering
\begin{tabular}{lcc}
\hline
 & SUA (RNS) & SOA (mSUGRA)   \\
\hline
$m_0$ & 5000 & 3500 \\
$m_{1/2}$  & 700& 500  \\
$A_0$ & -8300 & -7000  \\
$\tan\beta$  & 10 & 10  \\
$\mu $ & 110 & 2598.1 \\
$m_A$ & 1000 & 4284.2 \\
$m_h$ & 125.0 & 125.0 \\
$m_{\tg}$ & 1790 & 1312 \\
$m_{\tu}$ & 5100 & 3612 \\
$m_{\tst_1}$ & 1220 & 669 \\
$m_{\tz_1}$ & 101 & 224.1 \\
\hline
$\Omega^{\rm std}_{\tz_1} h^2$ & 0.008 & 6.8 \\
$\sigma^{\rm SI}(\tz_1 p)$ pb & $8.4\times 10^{-9}$ & $1.6\times 10^{-12}$ \\
\hline
\end{tabular}
\caption{Masses and parameters in~GeV units for two benchmark points
computed with Isajet 7.83 and using $m_t=173.2$ GeV.
}
\label{tab:bm}
\end{table}

The SUA point comes from {\it radiatively-driven natural SUSY}~\cite{rns}
with parameters from the 2-parameter non-universal Higgs model NUHM2
\be
\mbox{$(m_0,\ m_{1/2},\ A_0,\ \tan\beta )$ = $(5000\ {\rm GeV},\ 700\ {\rm GeV},\ -8300\ {\rm GeV},\ 10)$} .
\ee
with input parameters $(\mu,\ m_A)=(110,\ 1000)$ GeV~\cite{postlhc8}.
We generate the SUSY model spectra with Isajet 7.83~\cite{isajet}.
As shown in Table~\ref{tab:bm}, with $m_{\tg}=1.79$ TeV and $m_{\tq}\simeq 5$ TeV, 
it is allowed by LHC8 constraints on sparticles.
It has $m_h=125$ GeV and a higgsino-like neutralino with mass $m_{\tz_1}=101$ GeV
and standard thermal abundance
 of $\Omega_{\tz_1}^{MSSM}h^2=0.008$, low by a factor $\sim 15$ from the
 measured dark matter density~\cite{wmap9,Ade:2013zuv}.
Some relevant parameters, masses and direct detection cross sections
are listed in Table~\ref{tab:bm}.
It has very low electroweak finetuning.

For the SOA case, we adopt the mSUGRA/CMSSM model with parameters
\be
 \mbox{$(m_0,\ m_{1/2},\ A_0,\ \tan\beta ,\ sign(\mu ))$ = $(3500\ {\rm GeV},\ 500\ {\rm GeV},\ -7000\ {\rm GeV},\ 10,\ +).$}
\ee
The SOA point  has $m_{\tg}=1.3$ TeV and $m_{\tq}\simeq 3.6$ TeV,
so it is just beyond current LHC8 sparticle search constraints.
It is also consistent with the LHC Higgs discovery since $m_h=125$ GeV.
The lightest neutralino is mainly bino-like with $m_{\tz_1}=224.1$ GeV, and the
standard neutralino thermal abundance  is found to be
$\Omega_{\tz_1}^{\rm MSSM}h^2=6.8$,
a factor of $\sim 57$ above the measured value.
Due to its large $\mu$ parameter, this point has very high electroweak finetuning~\cite{sug_ft}.

\begin{figure}
\begin{center}
\includegraphics[height=9cm]{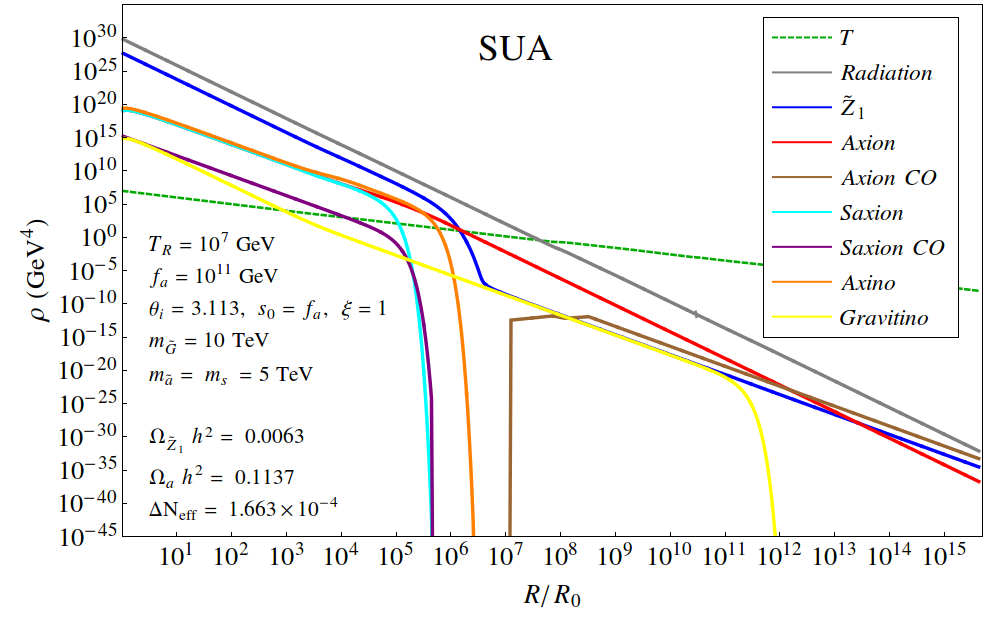}
\caption{Evolution of various energy densities vs. scale factor $R/R_0$
for the SUA benchmark case with $\xi=1$ and other parameters as indicated
in the figure.
\label{fig:rho1}}
\end{center}
\end{figure}
In Fig.~\ref{fig:rho1}, we show the solution of the Boltzmann equations
for the SUA point with $T_R=10^7$ GeV,
$f_a=10^{11}$ GeV, $m_{\tG}=10$ TeV, $m_{\ta}=m_s=5$ TeV, $\theta_s=1$, $\xi =1$ and
$\theta_i=3.11$. We present the evolution of the energy densities of axions and saxions  
(both CO- and thermally produced), axinos, neutralinos and gravitinos as a function of
the scale factor of the universe $R/R_0$, where $R_0$ is the scale factor at $T=T_R$.
For this parameter set, the final neutralino abundance is $\Omega_{\tz_1}h^2=0.0063$
whilst the axion abundance is $\Omega_ah^2=0.1137$, resulting in a total dark
matter relic abundance within the measured value.\footnote{
The standard thermal abundance of neutralinos calculated from our coupled
Boltzmann code is slightly below the IsaReD output due to our fit of
the IsaReD $\langle\sigma v\rangle (T)$ function.}
We see that at $T=T_R$ (where $R/R_0\equiv 1$) the
universe is radiation-dominated with smaller abundances of
neutralinos, axions, axinos and saxions, and even smaller abundances of CO-produced saxions and TP gravitinos. 
The CO-produced
saxions evolve as a non-relativistic matter fluid and so their density diverges from the relativistic
gravitino abundance as $R$ increases. Both TP- and CO- populations of saxions
begin to decay around $R/R_0\sim 10^5$, at temperatures ($T\sim 10^2$ GeV)
well below their masses. Somewhat later, but still before neutralino freeze-out, the axino population decays.
Since these decays happen before neutralino freeze-out, the TP-neutralino population is unaffected.
The axion mass turns on around $T\sim 1$ GeV so that the axion field 
begins to oscillate around $R/R_0\simeq 2\times 10^7$. The CO-produced axion field evolves
as CDM and ultimately dominates the universe at a value of $R/R_0$ somewhat off the plot.
The behavior of the DFSZ axinos and saxions-- 
in that they tend to decay before neutralino freeze-out-- 
is typical of this model for the lower range of $f_a\alt 10^{12}$
GeV with TeV-scale values of $m_{\ta}$ and $m_s$~\cite{dfsz1,ltr}.

Finally, gravitinos are long-lived and decay well after the neutralino
freeze-out, at $T \sim {\cal O}(100)$ keV. However, for $T_R = 10^7$ GeV,
gravitinos typically have a small number density and contribute marginally
to the final neutralino relic abundance. Also-- due to their small energy 
density-- the gravitino decays do not have any significant impact on big bang
nucleosynthesis.

In the following subsections, we compute the
neutralino and axion relic abundances for the two benchmark points through the
numerical integration of the Boltzmann equations presented in Sec.~\ref{sec:boltz}.
In order to be as general as possible, we will scan over the following SUSY DFSZ parameters:
\bea
 10^9 \mbox{ GeV} < & f_a & < 10^{16} \mbox{ GeV}, \nonumber \\
 0.4 \mbox{ TeV} < & m_{\ta} & < 20 \mbox{ TeV},\\
 0.4 \mbox{ TeV} < & m_s & < 20 \mbox{ TeV}. \nonumber \label{eq:scan}
\eea
For simplicity, we will fix the initial saxion field strength at $s_i=f_a$
($\theta_s\equiv s_i/f_a =1$) with $m_{\tG}=10$ TeV.
Unlike the SUSY KSVZ model, the bulk of our results do not strongly depend on
the re-heat temperature ($T_R$) since the axion, axino and saxion TP rates are
independent of this quantity.
Nonetheless, the gravitino thermal abundance is proportional to $T_R$
and since gravitinos are long-lived they may affect BBN if $T_R$ is
sufficiently large.
In order to avoid the BBN constraints on gravitinos, we choose $T_R=10^7$ GeV,
which results in a sufficiently small (would-be) gravitino abundance. 
As a result, gravitinos typically do not contribute significantly to the neutralino
abundance, as discussed above.

For each of the SUA and SOA benchmark points, we consider two different cases: $\xi=0$ and $\xi=1$.
As we can conclude from Eq.~\ref{eq:xicoup}, saxion decays into axions and axinos are turned off if $\xi=0$ whereas $s\to aa$ and $s\to\ta\ta$ decays are allowed for  $\xi=1$.

\subsection{Mixed axion/higgsino dark matter: SUA with $\xi=0$}
\label{sec:suaxi0}

In this section, we will examine the SUA SUSY benchmark assuming no
direct coupling between saxions and axions/axinos (see Eq.~\ref{eq:xicoup}),
which corresponds to $\xi =0$.  For each parameter set which yields an allowable value of $\Omega_{\tz_1}h^2<0.12$, we will
adjust the initial axion misalignment angle $\theta_i$ such that $\Omega_{\tz_1}h^2+\Omega_ah^2=0.12$, 
{\it i.e.} the summed CDM abundance saturates the measured value by 
adjusting the initial axion field strength parameter $\theta_i$.

\begin{figure}
\begin{center}
\includegraphics[height=6.9cm]{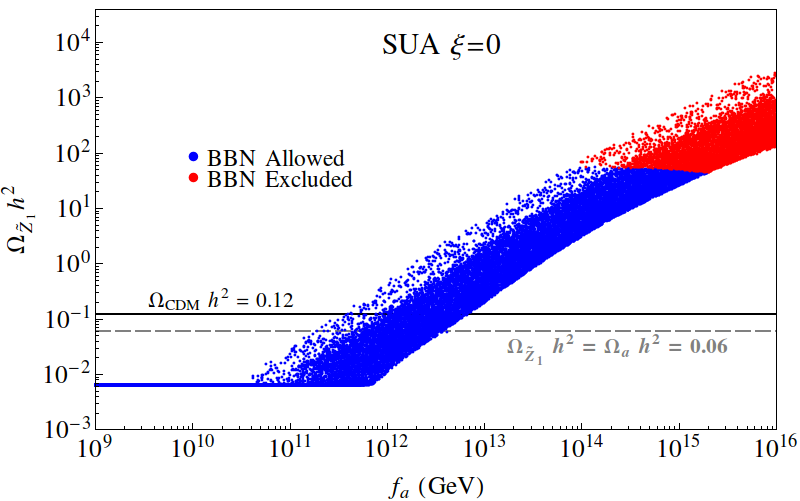}
\includegraphics[height=6.9cm]{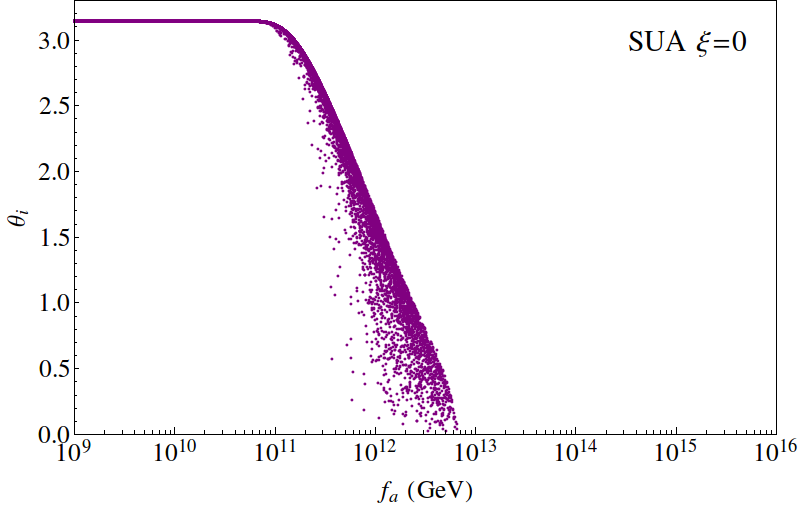}
\caption{In {\it a}) we plot the neutralino relic density from a scan over
SUSY DFSZ parameter space for the SUA benchmark case with $\xi=0$.
The grey dashed line shows the points where DM consists of 50\% axions and
50\% neutralinos.
In {\it b}), we plot the misalignment angle $\theta_i$ needed to saturate the dark matter relic density $\Omega_{\tz_1 a}h^2=0.12$.
\label{fig:sua0}}
\end{center}
\end{figure}
Our first results are shown in Fig.~\ref{fig:sua0}{\it a} where we plot 
$\Omega_{\tz_1}h^2$ vs. $f_a$ for a scan over the parameter space defined in
Eq.~\ref{eq:scan}.
Since for large $f_a$ values, saxions and axinos may decay during BBN,
we apply the BBN constraints using the bounds from Jedamzik~\cite{jedamzik} 
with extrapolations for intermediate values of
$m_X$ other than those shown in his plots. These constraints depend on the
lifetime of the decaying state, its energy density before decaying
and the fraction of energy injected as hadrons or color-charged states
($R_{h}$).
In the DFSZ scenario the dominant decays of saxions are into neutralinos, 
charginos, Higgs states or gauge bosons. 
Also, axinos decay into neutralinos or charginos plus gauge bosons or Higgs states. 
Thus the branching ratio for  $s\to\mbox{hadrons}$ must be similar to
$Br(W/Z\to\mbox{quarks})$ or $Br(\mbox{Higgs}\to\mbox{quarks})$, resulting
in $R_h \sim 1$. So we conservatively take $R_{h} = 1$ for saxion and axino decays. 
In Fig.~\ref{fig:sua0}{\it a} the red points violate BBN bounds on late-decaying
neutral relics, while the blue points are BBN safe. The points
below the solid gray line at 0.12 are DM-allowed, 
whilst those above the line overproduce neutralinos and so would be ruled out. 
The dashed gray line denotes the level of equal axion-neutralino DM densities: 
each at 50\% of the measured abundance.
Since, as previously discussed, the thermal production of axions gives
a negligible contribution to $\Delta N_{\rm eff}$ and, for  $\xi
=0$, there is no axion injection from saxion decays, dark radiation constraints
are always satisfied in this case.


For low values of $f_a\sim 10^9-10^{10}$ GeV, we see that $\Omega_{\tz_1}h^2$ takes on its standard
thermal value listed in Table~\ref{tab:bm}. 
This is because with such a small value of $f_a$, the
axino and saxion couplings to matter are sufficiently strong that they always decay before neutralino 
freeze-out. This behavior was also shown in Ref's~\cite{ltr,dfsz1} using semi-analytic calculations.
In this region, we expect mainly axion CDM with $\sim 5-10\%$ contribution of higgsino-like WIMPs~\cite{ltr}.
As $f_a$ increases, then saxions and axinos decay more slowly, and often after neutralino 
freeze-out. 
The late decays of saxions and axinos increases the neutralino density.
If the injection of neutralinos from saxion/axino decays is
sufficiently large, the `supersaturated' decay-produced neutralinos re-annihilate,
reducing their density. Although re-annihilation can reduce the neutralino
density by orders of magnitude, its final value is always larger than the
freeze-out density in the standard MSSM cosmology~\cite{ckls}.

As $f_a$ increases, the thermal production of axinos and saxions decreases,
while the density of CO-produced saxions increases (since we take $\theta_s = s_0/f_a =1$).
For $f_a \lesssim 10^{12}$ GeV, axinos and saxions are mostly thermally produced
and $\Omega_{\tz_1}h^2$ rises steadily with $f_a$ mainly due
to the increase of axino and saxion lifetimes, resulting in a late injection of
neutralinos well after their freeze-out. On the other hand, for 
$f_a \gtrsim 5\times10^{12}$ GeV, the thermal production of axions and axinos
becomes suppressed and the main contribution to the neutralino abundance comes from CO-produced
saxions and their decay. As seen in Fig.~\ref{fig:sua0}, once axinos and
saxions start to decay after the neutralino freeze-out ($f_a \gtrsim 5
\times 10^{10}$ GeV), $\Omega_{\tz_1}h^2$ always increases with $f_a$: this is
due to the increase in saxion and axino lifetimes and also due to the increase in rate of 
CO-produced saxions.
By the time $f_a$ exceeds $10^{13}$ GeV, then always too much
neutralino CDM is produced and the models are excluded. BBN constraints do not kick in until
$f_a$ exceeds $\sim 10^{14}$ GeV.
For a given $f_a$ value, the minimum value of $\Omega_{\tz_1}h^2$ seen
in Fig.~\ref{fig:sua0} happens for the largest
saxion/axino masses considered in our scan (20 TeV). This is simply due to the
fact that the lifetime decreases with the saxion/axino mass, resulting in
earlier decays. As a result, neutralinos are injected earlier on and can
re-annihilate more efficiently, since their annihilation rate increases with
temperature. Hence, an increase in the axino/saxion mass usually implies a
decrease in the neutralino relic abundance (for a fixed $f_a$ value).

In Fig.~\ref{fig:sua0}{\it b}, we show the value of the axion misalignment angle $\theta_i$ 
which is needed to obtain $\Omega_{\tz_1}h^2+\Omega_ah^2=0.12$. For 
low $f_a$ values ($\sim 10^9-10^{11}$ GeV), rather large values
of $\theta_i\sim\pi$ are required to bolster the axion abundance into the range
of the measured CDM density. For values of $f_a\sim 10^{11}-10^{12}$ GeV, then (perhaps more natural) 
values of $\theta_i\sim 2$ are required. For $f_a\agt4\times10^{12}$ GeV, axions tend to get overproduced by
CO-production and so a small value of $\theta_i\alt 0.5$ is required for suppression.
For even higher $f_a$ values, too many neutralinos are produced, so the models are all excluded. 


\subsection{Mixed axion/higgsino dark matter: SUA with $\xi=1$}
\label{sec:suaxi1}

We now discuss the main changes in the results of Fig.~\ref{fig:sua0} if we
consider a non-vanishing saxion-axion/axino coupling. For simplicity, we take
$\xi = 1$ where $\xi$ is defined in Eq.~\ref{eq:xicoup}.
In this case saxions can directly decay to axions and axinos (if $m_s >
2m_{\ta}$). The $s \to aa$ decay usually dominates over the other
decays~\cite{dfsz1}, suppressing $BR(s \to \ldots \to \tz_1 \tz_1)$ 
and significantly reducing the neutralino injection from saxion decays.
As a result, the neutralino relic abundance is usually smaller (for the same
choice of PQ parameters) than the $\xi = 0$ case.
Furthermore, the saxion lifetime is reduced (due to the large $s \to aa$ width)
and saxions tend to decay earlier when compared to the $\xi = 1$ case. 

\begin{figure}
\begin{center}
\includegraphics[height=6.9cm]{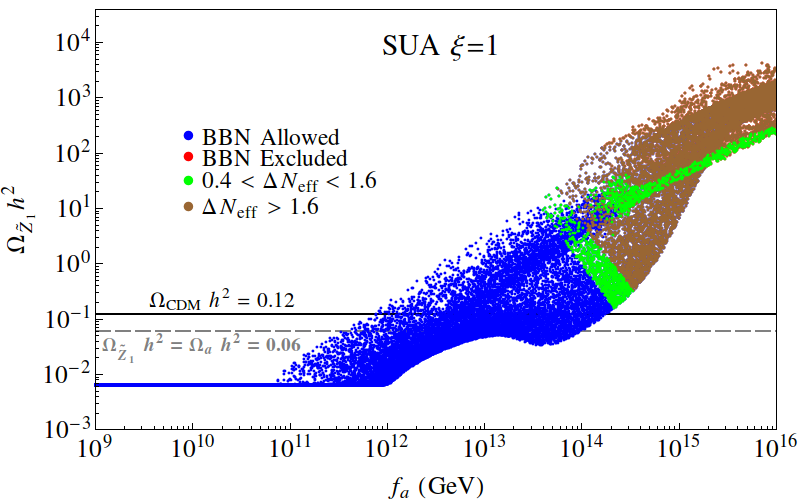}
\includegraphics[height=6.9cm]{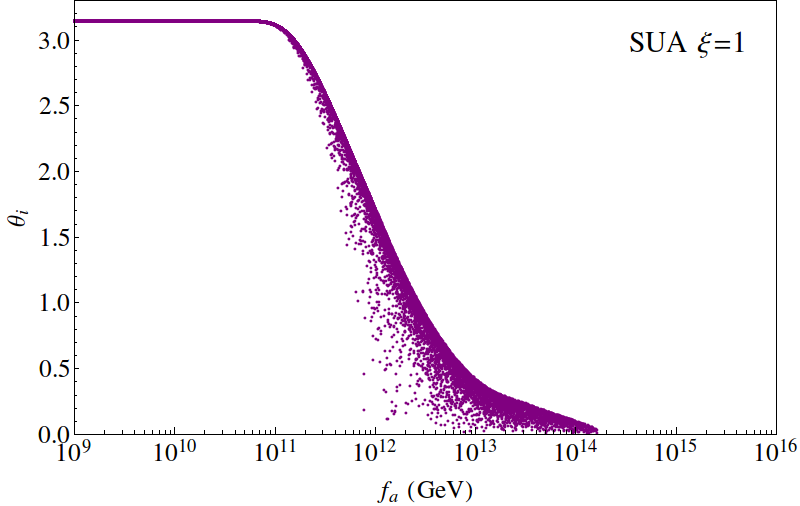}
\caption{In {\it a}) we plot the neutralino relic density from a scan over
SUSY DFSZ parameter space for the SUA benchmark case with $\xi=1$.
The grey dashed line shows the points where DM consists of 50\% axions and
50\% neutralinos.
The red BBN-forbidden points occur at $f_a\agt 10^{14}$ GeV and are covered over by the brown $\Delta N_{eff}>1.6$
coloration.
In {\it b}), we plot the misalignment angle $\theta_i$ needed to 
saturate the dark matter relic density $\Omega_{\tz_1 a}h^2=0.12$.
\label{fig:sua1}}
\end{center}
\end{figure}
In Fig.~\ref{fig:sua1}{\it a}, we once again show $\Omega_{\tz_1}h^2$ vs.
$f_a$ for the SUA SUSY benchmark but now for $\xi=1$. 
As just discussed, in this case the saxion lifetime is reduced,
so the region of $f_a$ where saxions/axinos always decay before
freeze-out is extended beyond the values generated for the $\xi=0$ case.
Since $BR(s \to \ldots \to \tz_1 \tz_1)$ is suppressed in the $\xi=1$ case,
saxions do not significantly contribute  to $\Omega_{\tz_1}h^2$ except when
$f_a \gtrsim 10^{14}$ GeV where CO-produced saxions have such large densities that--
even though their branching ratio to neutralinos is at the 0.1\% level-- 
their decay still enhances the neutralino relic density. 
For $10^{11} \mbox{ GeV} \lesssim f_a \lesssim 10^{14}$ GeV however, 
$\Omega_{\tz_1}h^2$ is dominated by the thermal axino contribution and 
the neutralino relic density increases with $f_a$, as in the
$\xi=0$ case. Once  $f_a \gtrsim 10^{13}$ GeV,
the thermal production of axinos becomes strongly suppressed and despite
decaying well after neutralino freeze-out, their contribution to
$\Omega_{\tz_1}h^2$ starts to decrease as $f_a$ increases. This is seen by the 
turn over of $\Omega_{\tz_1}h^2$ around $f_a \sim 10^{13}$ GeV.
As $f_a$ increases past $10^{14}$ GeV, CO saxions start to contribute to the
neutralino relic density, which once again rises with $f_a$.

Another important difference in the $\xi=1$ case is the large injection of
relativistic axions from saxion decays. For large values of $f_a$, where the
density of CO saxions is enhanced, the injected axions have a non-negligible
contribution to $\Delta N_{\rm eff}$. In particular, for $f_a \gtrsim 10^{14}$ GeV,
CO saxion decays produce too much dark radiation, so this region (shown by
brown points in Fig.~\ref{fig:sua1}{\it a}) is excluded by the CMB constraints
on dark radiation ($\Delta N_{\rm eff} < 1.6$).\footnote{There is some  tension
between the current Planck, WMAP and BBN values for $\Delta N_{\rm eff}$. Hence we
take this number as a conservative bound, as discussed in
Ref.~\cite{bbl}.} These points are also excluded by overproduction of neutralinos and violation of BBN bounds.
We also show as green points the cases where
$\Delta N_{\rm eff}\sim 0.4-1.6$ which could explain a possible excess of dark
radiation suggested by the combined WMAP9 result. However these points are
already excluded by overproduction of dark matter.

Finally, in Fig.~\ref{fig:sua1}{\it b}, we again plot the value of $\theta_i$
which is needed by axions so that one matches the measured abundance of CDM, as 
described in the previous section. Once again, 
at low $f_a$, $|\theta_i |\sim \pi $ is required, while
for high $f_a$ values ($\agt 10^{13}$ GeV), low $|\theta_i |$ is
required in order to suppress axion CO-production.
Furthermore, since $\Omega_{\tz_1}h^2$ is usually smaller in the $\xi=1$
case for the same $f_a$ values (when compared to $\xi=0$), the CO axion contribution to DM can be
larger and higher values of $\theta_i$ are usually allowed, as seen in
Fig.~\ref{fig:sua1}{\it b}.


\subsection{Mixed axion/bino dark matter: SOA with $\xi=0$}
\label{sec:soaxi0}

In this Section, we turn to the SUSY benchmark SOA, which features a bino-like
LSP with a standard thermal overabundance $\Omega_{\tz_1}h^2=6.8$, {\it i.e.}
too much dark matter by a factor 57! The SUSY $\mu$ parameter has a value of
$\mu =2598$ GeV so this model would be considered fine-tuned in the electroweak sector.
However, the large $\mu$-parameter also bolsters the saxion and axino decay rates which are proportional to
some power of $\mu$ ($\mu^2$ or $\mu^4$) in the SUSY DFSZ model~\cite{dfsz1}.

\begin{figure}
\begin{center}
\includegraphics[height=6.9cm]{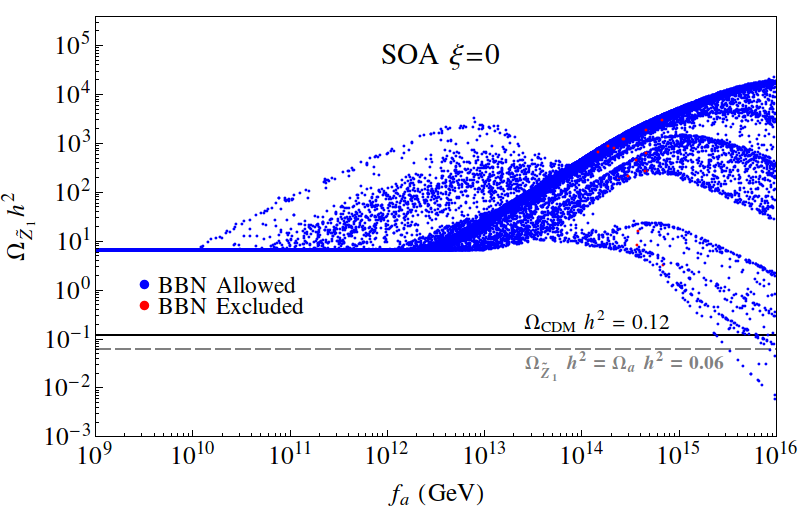}
\caption{We plot the neutralino relic density from a scan over
SUSY DFSZ parameter space for the SOA benchmark case with $\xi=0$.
The grey dashed line shows the points where DM consists of 50\% axions and
50\% neutralinos.
\label{fig:soa0}}
\end{center}
\end{figure}
In Fig.~\ref{fig:soa0}, we show the coupled Boltzmann calculation of
$\Omega_{\tz_1}h^2$ as a function of $f_a$ for the SOA benchmark with $\xi=0$.
At low $f_a\sim 10^9-10^{10}$ GeV, axinos and saxions decay before neutralino freeze-out, so the
model remains excluded due to overproduction of dark matter. As $f_a$ increases,
neutralinos are only produced at higher and higher rates as their population is
bolstered by late time axino and saxion decay, as already observed for the SUA
case. In this region, the highest $\Omega_{\tz_1}h^2$ values for
a fixed $f_a$ are obtained for the smallest $m_s$, $m_{\ta}$ values, 
since these correspond to the longest lifetimes.
However, once $f_a\gtrsim 10^{14}$ GeV, a subset of points present the opposite
behavior and the neutralino relic abundance actually {\it decreases} with
$f_a$. This region of parameter space corresponds to small saxion masses, $m_s
\lesssim 2 m_{\tz_1}$, so the decay to neutralinos is kinematically forbidden.
As a result (since $\xi=0$, saxions do not decay to axions) the only effect
of saxion decays is to inject entropy in the early universe. For $f_a\gtrsim
10^{15}$ GeV, there is a huge rate for saxion production via coherent
oscillations and the entropy injection from saxion decays can reduce the
neutralino density, resulting in DM-allowed models with
$\Omega_{\tz_1}h^2<0.12$. We discuss these cases in detail in
Sec.~\ref{sec:soasax}.
We also point out that in the SUA model or in the SUSY KSVZ case, such large
$f_a$ values imply very long-lived saxions, with lifetimes of the order of
$\mathcal{O}(10\,s)$ or greater. As a result, all
the solutions with large entropy injection in the SUA case are excluded by BBN
contraints.\footnote{We stress however, that this result relies on the assumption that the saxion initial field value is given
by the PQ breaking scale ($\theta_s = s_0/f_a = 1$). As shown in
Ref.~\cite{bls}, in the KSVZ case the neutralino relic abundance can be
suppressed if one takes $s_0 \gg f_a$ or $\theta_s \gg 1$.}
However, for the SOA case, the large $\mu$ value enhances the saxion  decay
rate to Higgs pairs and vector bosons and even at such high $f_a$ values,
saxions can still decay before BBN starts.
Very few points do succumb to BBN constraints (denoted by red points) 
but  these are also excluded due to an overabundance of neutralinos.
In Fig.~\ref{fig:soa0} we also see that in the large $f_a$ region there is a
visible gap (for a fixed $f_a$ value) between the branch with a suppression of
$\Omega_{\tz_1}h^2$ and the one with an enhanced value of $\Omega_{\tz_1}h^2$.
The lower branch (with $\Omega_{\tz_1}h^2 \lesssim 20$) corresponds to 
points with low saxion masses, where $BR(s \to \ldots \tz_1 \tz_1) \ll 1$, so
saxion decays mostly dilute the neutralino relic density. Once $m_s > 2 m_{\tilde{t}_1}$,
the $s \to \tilde{t}_1 \bar{\tilde{t}}_1$ channel becomes kinematically allowed and
there is a sudden increase in $\Omega_{\tz_1}h^2$, resulting in the gap seen in
Fig.~\ref{fig:soa0}.
Finally, since $\xi =0$, axions are only thermally produced resulting in a
negligible contribution to $\Delta N_{\rm eff}$, so dark radiation constraints are inapplicable in this case.

%
\subsection{Mixed axion/bino dark matter: SOA with $\xi=1$}
\label{sec:soaxi1}

\begin{figure}
\begin{center}
\includegraphics[height=6.9cm]{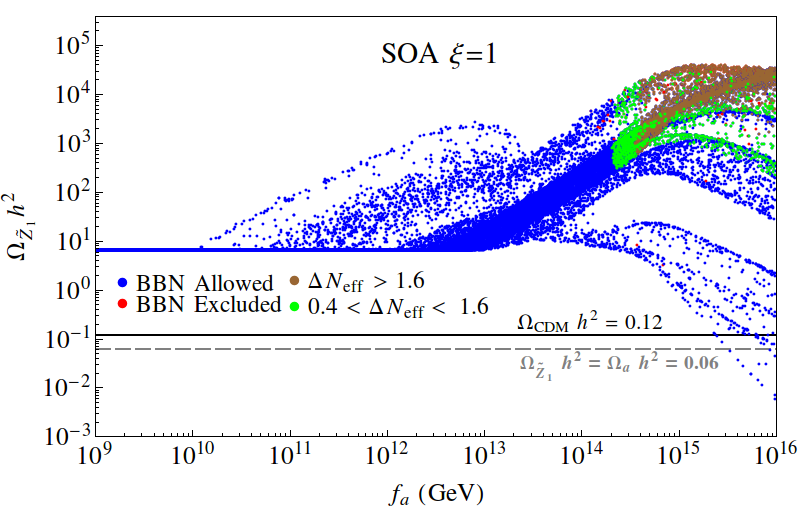}
\caption{We plot the neutralino relic density from a scan over
SUSY DFSZ parameter space for the SOA benchmark case with $\xi=1$.
The grey dashed line shows the points where DM consists of 50\% axions and
50\% neutralinos.
\label{fig:soa1}}
\end{center}
\end{figure}
In Fig.~\ref{fig:soa1} we plot $\Omega_{\tz_1}h^2$ vs. $f_a$ for the SOA SUSY benchmark but with $\xi=1$.
Unlike the SUA case, decays to axions are not
always dominant, since $\Gamma(s \to aa)
\sim m_s^3/f_a^2 $, while $\Gamma (s\to VV,\ hh)\sim \mu^4/(m_s f_a^2)$. Hence
saxions dominantly decay to gauge bosons/higgses, except for $m_s \gg \mu$.
The low $f_a$ behavior of $\Omega_{\tz_1}h^2$ is much the same as in the $\xi=0$
case: the neutralino abundance is only bolstered to even higher values
and thus remains excluded by overproduction of WIMPs.
As in the SOA $\xi=0$ case, there again exists a set of
points with $f_a\agt 10^{15}$ GeV and with $m_s \lesssim 2m_{\tz_1}$ which is
allowed by all constraints.
This is possible in the $\xi=1$ case, since, for $m_s \ll \mu$, saxions mainly
decay to higgses and gauge bosons, thus injecting enough entropy to dilute
$\Omega_{\tz_1}h^2$. Points with $m_s \gg \mu$, however, have $BR(s \to aa)
\simeq 1$, resulting in a large injection of relativistic axions and a
suppression of entropy injection.
In this case many models start to become excluded by overproduction of dark
radiation (brown points) while some also have $\Delta N_{\rm eff}\sim 0.4-1.6$:
these points could explain a possible  excess of dark radiation except 
that they also always overproduce neutralino dark matter. 
Thus, we see that the SUSY DFSZ model with large $\mu$ and either small or large $\xi$ 
along with small $m_s$ is able to reconcile the expected value of Peccei-Quinn scale~\cite{bl} 
from string theory~\cite{dine2,witten} (where $f_a$ is expected $\sim m_{\rm GUT}$) 
with dark matter abundance, dark radiation and BBN constraints.

\subsection{Mixed axion/bino dark matter with a light saxion}
\label{sec:soasax}
\begin{figure}
\begin{center}
\includegraphics[height=9cm]{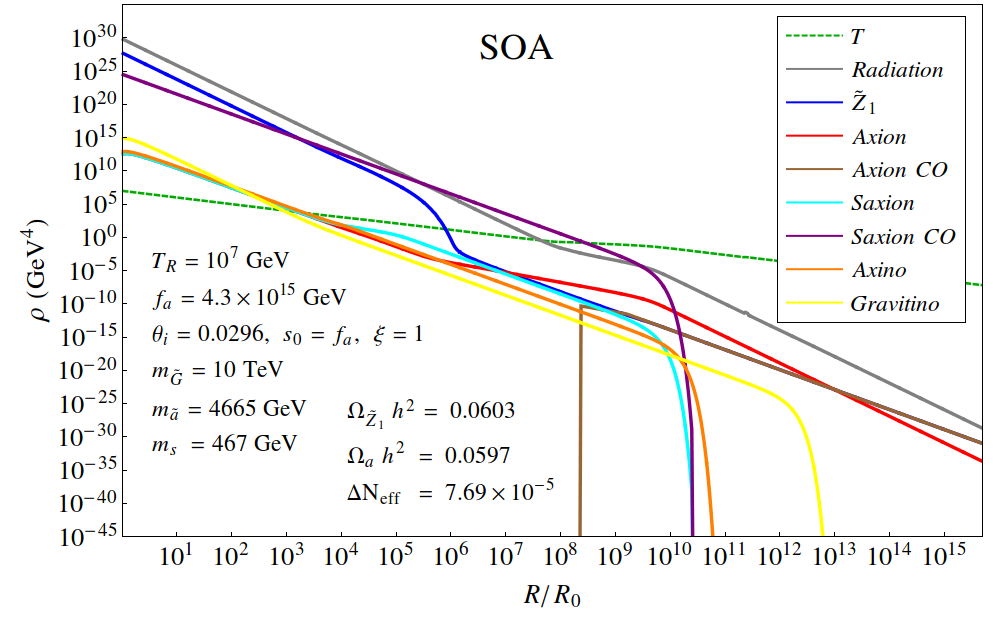}
\caption{Evolution of various energy densities vs. scale factor $R/R_0$ 
for the SOA benchmark case with $\xi=1$.
\label{fig:rho2}}
\end{center}
\end{figure}
As discussed in the previous sections, the neutralino relic abundance can only
be suppressed with respect to its MSSM value if $m_s \lesssim 2 m_{\tz_1}$ and
$f_a \gtrsim 10^{15}$ GeV. Here we discuss in detail this case, since it
represents the only possibility to reconcile the SOA dark matter
scenario with the measured DM abundance.
A specific example is shown in Fig.~\ref{fig:rho2}, where the evolution of the
energy density of various species as a function of the universe scale factor is
presented for $f_a = 4.3 \times 10^{15}$ GeV, $m_s = 467$ GeV and
$m_{\ta} = 4.67$ TeV. For this choice of parameters, the neutralino relic
abundance is highly suppressed ($\Omega_{\tz_1} h^2=0.06$) but does comprise 
50\% of the total DM abundance. 
The remainding 50\% is composed of axions although these require a somewhat
small value of the axion mis-alignment angle ($\theta_i = 0.03$) in order to suppress 
the CO axion production.
From Fig.~\ref{fig:rho2} we see that the CO-produced saxion energy density
dominates over the radiation energy density at $R/R_0\sim 10^6$ and decays at $R/R_0\sim10^{10}$, 
so that the universe is saxion-dominated during this period.
In this case, saxions dominantly decay into SM particles, since the rate for
saxion $\to$ neutralinos is highly suppressed by the kinematic phase factor 
($BR(s\to\widetilde{Z}_1\widetilde{Z}_1)\sim10^{-8}$ at this point).
Therefore, a huge amount of entropy is produced as we can see from the radiation
curve (grey), while the neutralino density (blue) is almost unaffected
by the saxion decay.
As a result, the final neutralino density is given by
$\Omega_{\widetilde{Z}_1}=0.06$ and this can be a viable model, even though
the PQ scale is very large.

\begin{figure}
\begin{center}
\includegraphics[height=6.9cm]{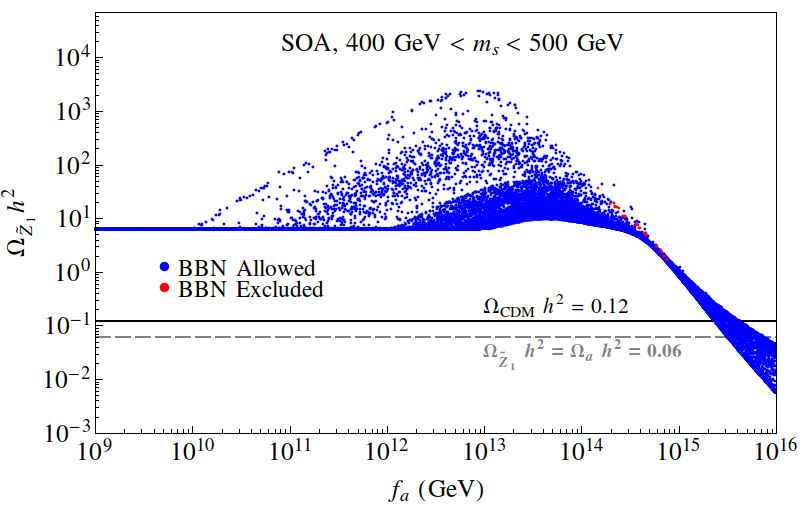}
\includegraphics[height=6.9cm]{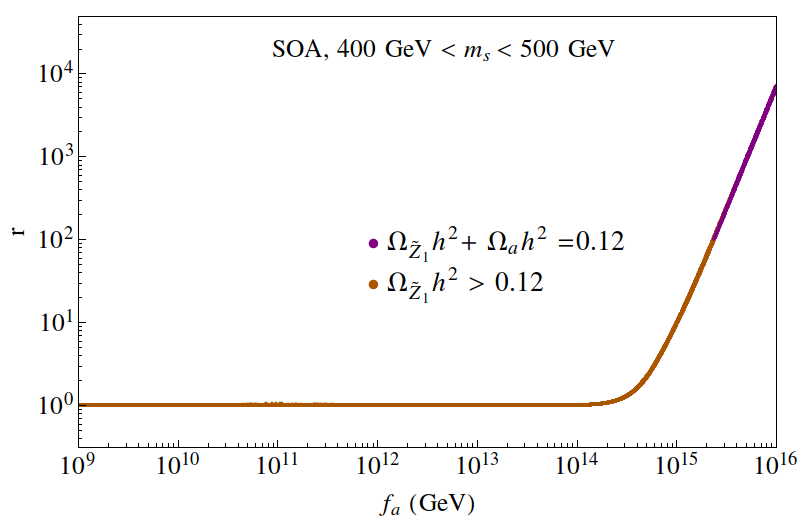}
\caption{In {\it a}), we plot the neutralino relic density vs $f_a$ 
for the scan over the SUSY DFSZ parameter space for the SOA benchmark case 
with $\xi= 0$ and 1, but with $m_s:400-500$ GeV.
The grey dashed line shows the points where DM consists of 50\% axions and
50\% neutralinos.
In {\it b}) we plot the dilution factor $r$ vs. $f_a$.
\label{fig:dilute}}
\end{center}
\end{figure}
In Fig.~\ref{fig:dilute}, we once again scan over the parameter space from
Eq.~\ref{eq:scan}, but now we focus on the light
saxion region, $400$ GeV$<m_s<500$ GeV, where the saxion decay to neutralinos
is kinematically suppressed or forbidden.
As already seen in  Figs.~\ref{fig:soa0} and \ref{fig:soa1}, in this case the
large $f_a$ region ($f_a \gtrsim 10^{15}$ GeV) can suppress $\Omega_{\tz_1} h^2$
to values below the observed DM abundance.
Fig.~\ref{fig:dilute}{\it b} shows how the entropy dilution factor ($r
\equiv S_f/S_0$) increases with $f_a$, reaching values as high as $10^4$, for
$f_a\sim10^{16}$ GeV.


We note here that one might wonder if the large $f_a\sim m_{\rm GUT}$ region of the SUA
model might be DM-allowed if we consider $m_s<200$ GeV so that saxion decay to SUSY particles
is dis-allowed and saxion decay leads only to entropy dilution. Aside from the fact that such
light values of $m_s$ leads to a large disparity between scalar soft breaking terms, in the
SUSY DFSZ model these points should all be BBN dis-allowed.

\section{Conclusion}
\label{sec:conclude}

In this paper, we have discussed the evaluation of the relevant Boltzmann
equations in the supersymmetrized DFSZ axion model.
This is a highly motivated scenario, since it provides
solutions to the gauge hierarchy and strong CP problems as well as a
solution to the SUSY $\mu$ problem while allowing for the Little Hierarchy $\mu \ll
m_{3/2}$ which is expected from combining naturalness considerations with LHC bounds 
on sparticle masses and the 125 GeV Higgs boson mass. 
In SUSY DFSZ, axinos and saxions tend to decay to vector bosons, Higgs states and higgsinos. 
Saxions may also decay into $aa$ or $\ta\ta$, depending on the value of the
saxion-axion/axino model dependent coupling $\xi$. 
The first of these leads to dark radiation while the second may enhance the
neutralino relic density.

In the SUSY DFSZ scenario, the decay widths of saxions and axinos are enhanced
for large $\mu$ values and their decays may happen at temperatures of the order
of their masses. Hence it is crucial to include the inverse decay processes in
the Boltzmann equations. Furthermore, since in the SUSY DFSZ case the thermal
production of saxions, axions and axinos happen through the freeze-in mechanism,
the production and decay processes may happen at similar time scales. 
In these cases, a precise calculation of the saxion and axino evolution is only
possible through the numerical integration of the Boltzmann equations.

Since most of the axion supermultiplet couplings in the SUSY DFSZ model are
proportional to $\mu$, we have presented results for two SUSY benchmark points:
1. a natural SUSY model labelled SUA with $\mu =110$ GeV and a higgsino LSP, and 
2. a mSUGRA/CMSSM  point (SOA) with $\mu = 2.6$ TeV and a bino-like LSP,
resulting in a standard thermal neutralino overabundance.
We found that, for the SUA benchmark with $\xi =0$, low $f_a\sim
10^9-10^{11}$ GeV tends to give mainly axion CDM with 5-10\%
higgsino-like WIMPs. For higher $f_a$ ($\sim 10^{11}-10^{12}$ GeV), the WIMP
density increases and might even dominate the DM abundance. For $f_a\agt 6\times10^{12}$ GeV, 
the model becomes excluded due to overproduction of WIMPs. For SUA with $\xi =1$
the contribution of $s\to aa$ hastens the saxion decay rate so that saxion decay occurs before neutralino freeze-out
over an even larger range of $f_a$. 
In this case, for sufficiently heavy saxions and axinos, $f_a\sim 10^9-10^{14}$
GeV is allowed by all constraints.
For even higher $f_a$ values ($f_a\agt 2\times10^{14}$ GeV), the model becomes excluded by
overproduction of WIMPs, overproduction of dark radiation and violation of BBN constraints.

For the SOA model, the presence of axions, saxions and axinos typically leads to
an enhancement of the neutralino relic abundance for almost the entire $f_a$
range, so such models typically remain excluded. The exception comes at
very large $f_a$ values ($\sim 10^{15}-10^{16}$ GeV) with small saxion masses,
$m_s\lesssim 2m_{\tz_1}$.
In this case, enormous entropy injection from CO-produced saxions along with
their decays to SM particles leads to entropy dilution of the WIMP relic
density whilst avoiding BBN and dark radiation constraints.
\begin{figure}
\begin{center}
\includegraphics[height=4.6cm]{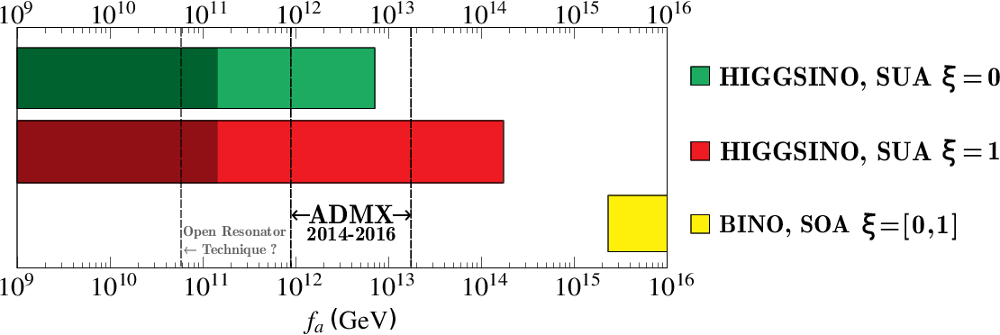}
\caption{Range of $f_a$ which is allowed in each PQMSSM scenario for the SUA and SOA
benchmark models. Shaded regions indicate the range of $f_a$ where $\theta_i>3$.
\label{fig:bar}}
\end{center}
\end{figure}

An overview of our results is presented in Fig.~\ref{fig:bar} where we show the allowed
range of $f_a$  as a bar for SUA and SOA models. We also denote the range of $f_a$ values which are expected to be probed in the next few years by the ADMX experiment~\cite{admx}. A possible ADMX technique of open resonators discussed in~\cite{Rybka:2014cya} may allow lower values of $f_a$ to be probed in the future.

For all allowed cases, we would ultimately expect both WIMP and axion
dark matter detection to occur.

\acknowledgments

We thank E. J. Chun for earlier collaboration on these topics and B. Altunkaynak for his assistance with supercomputing.
The computing for this project was performed at the OU Supercomputing Center for Education \& Research (OSCER) at the University of Oklahoma (OU).

\end{document}